\documentclass[12pt,preprint]{aastex}  

\slugcomment{Accepted by the Astronomical Journal}   
   
\begin{document}   
   
\def\ang{{\rm\AA}}   
   
\def\n{\noindent}   
   
\def\ph{\phantom}   
   
\newcommand{\myfigure}[4]{   
        \begin{figure*}   
        \setbox100=\hbox{   
        \epsfxsize=#1 cm   
        \epsfbox{#2.ps}}   
        \centerline{\hbox{\box100}}   
        \caption{#3}   
        \label{#4}   
        \end{figure*}   
}   
   
\title{The Ultraviolet Spectra of the Weak Emission Line Central     
Stars of Planetary Nebulae}   
   
\author{W. L. F. Marcolino\altaffilmark{1,2}}   
\email{wagner@on.br}   
   
\author{F. X. de Ara\'ujo\altaffilmark{1}}   
\email{araujo@on.br}      

\author{H. B. M. Junior\altaffilmark{1,3}}   
\email{helson@on.br}   
   
\and

\author{E. S. Duarte\altaffilmark{3}}   
\email{esduarte@cefeteq.br}    
   
\altaffiltext{1}{Observat\'orio Nacional/MCT, Rua Gen. Jos\'e Cristino 77,   
  20921-400, Rio de Janeiro, Brazil.}   
  
\altaffiltext{2}{Laboratoire d'Astrophysique de Marseille, Traverse du Siphon-Bp 8, 13376 Marseille Cedex 12, France.}
 
\altaffiltext{3}{Centro Federal de Educa\c c\~ao Tecnol\'ogica de Qu\'imica de 
Nil\'opolis, 26530-060, Rio de Janeiro, Brazil.}    
   
%\bigskip   
   
\begin{abstract}   
   
The ultraviolet spectra of all ``weak emission line central stars of  
planetary nebulae'' (WELS) with available {\it IUE} data are presented and discussed.  
We performed line identifications, equivalent width and flux measurements for
several features in their spectra. We found that the WELS can be divided in three    
different groups regarding their UV: i) Strong P-Cygni profiles       
(mainly in \ion{C}{4} $\lambda 1549$); ii) Weak P-Cygni features    
and iii) Absence of P-Cygni profiles. The last group encompasses    
stars with a featureless UV spectrum or with intense emission lines and a    
weak continuum, which are most likely of nebular origin. We have measured wind    
terminal velocities for all objects presenting P-Cygni profiles in    
\ion{N}{5} $\lambda 1238$  and/or \ion{C}{4} $\lambda 1549$.    
The results obtained were compared to the UV data of the two prototype stars of    
the [WC]-PG 1159 class, namely, A30 and A78. For WELS presenting P-Cygnis, 
most of the terminal velocities fall in the range $\sim 1000-1500$ km 
s$^{-1}$, while [WC]-PG 1159 stars possess much higher values, of  
$\sim$3000 km s$^{-1}$. The [WC]-PG1159 stars are characterized by  
intense, simultaneous P-Cygni emissions in the $\sim 1150-2000$\AA\,  
interval of \ion{N}{5} $\lambda 1238$, \ion{O}{5} $\lambda 1371$ and  
\ion{C}{4} $\lambda 1549$. In contrast, we found that  
\ion{O}{5} $\lambda 1371$ is very weak or absent in the WELS  
spectra. On the basis of the ultraviolet spectra alone, our findings  
indicate that [WC]-PG 1159 stars are distinct from the WELS, contrary  
to previous claims in the literature.    
   
\end{abstract}   
   
   
%%%%%%%%%%%%%%%%%%%%%%%%%%%%%%%%%%%%%%%%%%%%%%%%%%%%%%%%%%%%%%%%%%%%%%%%   
   
\section{INTRODUCTION}   
   
Central stars of planetary nebulae (CSPN) that are hydrogen deficient    
are generally divided in three main groups: [WR], PG 1159 and [WC]-PG 1159    
stars. The first one present spectra with strong and broad emission lines  
mainly from He, C and O, similar to the Wolf-Rayet stars (WR) of Population I.    
Depending on the ionization stages of the elements dominating the atmosphere,    
[WR] stars are subdivided in [WCL], [WCE] and [WO] (Crowther et al. 1998;    
Acker \& Neiner 2003). On the other hand, PG 1159 stars are quite distinct  
objects. They are pre-white dwarfs and show mainly absorption lines of \ion{He}{2} and  
\ion{C}{4} in their spectra (Werner et al. 1997). Only a handful of   
PG 1159 stars are known to possess wind features (Koesterke et al. 1998;   
Koesterke \& Werner 1998). The [WC]-PG 1159 group present strong P-Cygni   
lines in the ultraviolet (e.g. \ion{N}{5} $\lambda 1238$ and \ion{C}{4}  
$\lambda 1549$) and resemble the PG 1159 stars in the optical. Three objects  
are considered  prototypes of this class: A 30, A 78 and Longmore 4 (at 
outburst).   
   
The origin and evolution of hydrogen deficient CSPN have been investigated    
from different point of views during the last decade. Evolutionary    
models are now able to provide a reasonable match to the observed    
chemical abundances, although it is still debated the role   
of binarity and different thermal pulse models (De Marco \& Soker 2002; Herwig  
2001). Nebular analyses as well as sophisticated non LTE atmosphere models have been 
also very useful to address several questions regarding these central stars:  
it is generally inferred that the groups above mentioned form  
the following evolutionary sequence: late type [WR] $\rightarrow$ early-type  
[WR] $\rightarrow$ [WC]-PG 1159 $\rightarrow$ PG 1159 $\rightarrow$ non DA   
white dwarfs (see e.g. Zijlstra et al. 1994; Pen\~a et al. 2001; Koesterke  
2001). However, this scenario has important issues unsolved, such as the C/He   
mass ratio and the exact position of [WC]-PG 1159 stars in the HR diagram   
(Hamann 1997; Marcolino et al. 2007). Moreover, the evolutionary status  
of the so-called ``weak emission line stars'' and their relation to the  
[WC]-PG 1159 stars is not at all clear. Let us discuss it now. 
  
In an extensive observational study, Tylenda et al. (1993) analyzed 
spectroscopy data of 77 hydrogen deficient CSPN.  
In their sample, 39 were classified as [WR] stars. The   
remaining objects were called ``weak emission line stars'' (WELS or [WELS]).   
According to these authors, the WELS show emission of \ion{C}{4} $\lambda 5805$    
(actually \ion{C}{4} $\lambda \lambda 5801, 12$) systematically    
weaker and narrower than in [WR] stars, and a feature in $\lambda 4650$,    
which is possibly a blend of \ion{N}{3}, \ion{C}{3} and \ion{C}{4} emissions.    
Moreover, \ion{C}{3} $\lambda 5696$ is very weak or absent.    
These optical characteristics were confirmed by Marcolino \& de Ara\'ujo  
(2003) with a homogeneous and higher resolution set of data.  
Interestingly, by comparing a large sample of WELS, [WC]-PG 1159 and    
PG 1159 spectra, Parthasarathy et al. (1998) claimed that    
the WELS and the [WC]-PG 1159 stars actually constitute the same class.   
However, this claim was based solely on comparisons of optical spectra and    
was not confirmed by further studies. Indeed, some authors    
argued that this assertion should be taken with caution until    
an analysis of a larger sample of objects is performed    
(e.g. Werner \& Herwig 2006).  
   
Undoubtedly, the ``weak emission line stars'' (WELS) constitute  
the least understood class of hydrogen deficient CSPN. An important point 
is that the WELS might not be descendants of the [WR] stars, as it is 
explicited in the [WR] $\rightarrow$ [WC]-PG 1159 (=WELS) $\rightarrow$ 
PG 1159 evolution. Pen\~a et al. (2003) for example, derived lower average nebular 
expansion velocities for WELS than for [WR] stars, while the 
contrary would be expected from the long-term action of a stellar 
wind on a planetary nebula during a [WR] $\rightarrow$ WELS transition. 
Furthermore, based on a kinematical study of a large sample of planetary nebulae, Gesicki et al. (2006)  
raised the interesting possibility that some WELS can be progenitors  
of the hottest [WR] stars, i.e., the [WO] group (WELS $\rightarrow$ [WO]).  
A major issue hinders the determination of the evolutionary status of  
the WELS: their physical parameters (e.g. $T_{eff}$, $v_\infty$  
and $\dot{M}$) and chemical abundances remain unknown. The properties  
of A 30, A 78, and Longmore 4 ([WC]-PG 1159 stars) are known  
(Koesterke 2001) but again, their identity as WELS is questionable.  
  
So far, most of previous studies involving WELS were done in the optical part of   
the spectrum. In fact, the very definition of a WELS is based in the optical   
features in $\lambda 4650$, \ion{C}{4} $\lambda 5805$ and \ion{C}{3} $\lambda 5696$  
(Tylenda et al. 1993). A few works in the ultraviolet including some WELS can   
be found in the literature. However, they often include stars of different 
spectral classes (e.g. Feibelman 2000), and/or are focused in the
determination of nebular properties (e.g. Adams \& Seaton 1982;
Pottasch et al. 2005), without special attention to the WELS and their 
evolutionary state. Motivated by this fact and considering the open questions above described,   
we investigate in this paper the UV spectra of all WELS with available 
{\it IUE} ({\it International Ultraviolet Explorer}) data.  
Our main aims are to understand their main UV characteristics; identify and 
measure the most intense spectral lines; measure wind terminal velocities from  
the available P-Cygni profiles; and finally, to compare the results to the  
data of the two prototype [WC]-PG 1159 stars: A30 and A78.  
 
The present paper is organized as follows: in Section 2 we present  
the observational data retrieved from the Multi-mission Archive at STScI  
(MAST); in Section 3 we discuss the main characteristics  
of the UV spectra of the WELS, and present line identifications  
of the most conspicuous features, as well as equivalent  
width and line flux measurements. In Section 4   
we empirically measure the terminal velocities for  
all objects presenting P-Cygni profiles in \ion{N}{5} $\lambda 1238$  
and/or \ion{C}{4} $\lambda 1549$, from low and high resolution data.  
Finally, in Section 5, we discuss and compare the results obtained for 
the WELS to the data of A 30 and A 78 (the two prototype [WC]-PG 1159 stars), 
and present the main conclusions of our work.  
  
%%%%%%%%%%%%%%%%%%%%%%%%%%%%%%%%%%%%%%%%%%%%%%%%%%%%%%%%%%%%%%%%%%%%%   
   
\section{OBSERVATIONAL DATA}   
 
We used public data from the {\it IUE} ({\it International Ultraviolet 
Explorer}) satellite, which are available at the Multi-mission Archive  
at STScI (MAST)\footnote{http://archive.stsci.edu/}. The total  
number of WELS currently known is about 50. Most of them are  
listed in the work of Tylenda et al. (1993) and Parthasarathy  
et al. (1998), and 15 were recently discovered in the  
direction of the Galactic bulge by G\'orny et al. (2004). Two objects 
considered as WELS in Parthasarathy et al. (1998) (Hen 2-86 and M 2-31)  
were found to be actually early type [WR] stars by Marcolino \& de Ara\'ujo  
(2003) on the basis of the equivalent width of the \ion{C}{4} $\lambda 5805$ 
line.  
 
We have searched for {\it IUE} data for all WELS known, and found spectra  
for only $\sim 40\%$ of the total WELS population. The list of  
SWP, LWP and LWR archives considered in our study, including the 
ones for A30 and A78 (the two [WC]-PG 1159 stars), is shown in Table \ref{objects}. 
The majority of the spectra disponible are of low resolution ($\sim 6$\AA). 
The high resolution spectra ($\sim 0.2$\AA) which were utilized are also 
shown in the table. Most of the observations were done using the {\it IUE} 
large aperture mode ($10'' \times 20''$). 
 
%%%%%%%%%%%%%%%%%%%%%%%%%%%%%%%%%%%%%%%%%%%%%%%%%%%%%%%%%%%%%%%%%%   
   
\section{THE ULTRAVIOLET SPECTRUM OF WELS}   
\label{uvwels} 
   
The ``weak emission line'' class of central stars is defined based on  
the optical spectrum. Tylenda et al. (1993) showed that the WELS are  
characterized mainly by a feature in $\lambda 4650$ (possibly a sum of   
\ion{N}{3}, \ion{C}{3} and \ion{C}{4}), and the \ion{C}{3} $\lambda 5696$ and  
\ion{C}{4} $\lambda 5805$ transitions. Because their \ion{C}{4} $\lambda 5805$ line 
is usually weak, a [WCL] spectral classification seems reasonable for  
these stars at a first glance. However, the \ion{C}{3} $\lambda 5696$ line  
is generally strong in the [WCL] class and it is weak or absent  
in the WELS. Conversely, because the \ion{C}{3} $\lambda 5696$ is  
weak or absent, a [WCE] classification seems plausible. However,  
the \ion{C}{4} $\lambda 5805$ in [WCE] stars is much stronger than  
in the WELS. Besides these characteristics, the WELS are also known  
to present some absorptions in the optical from lines such as  
\ion{He}{2} $\lambda 4541$ and $\lambda 5412$ (Parthasarathy et al. 1998;  
Marcolino \& de Ara\'ujo 2003), indicating a stellar wind not  
so dense as it is the case of the [WR] stars, where the photospheric  
part is completely hidden. 

Although we can say that the main characteristics of the WELS  
are known in the optical, their ultraviolet spectra were never  
discussed before. As we have mentioned, although some works  
in the UV involved some WELS, no attention was given to their 
evolutionary status or relation to the [WC]-PG 1159 stars. 
The present paper addresses directly this issue. Instead of 
presenting a homogeneous set of features, we found that the WELS 
can be divided in three main groups regarding their ultraviolet spectra:  
 
\begin{enumerate} 
 
\item presence of a strong P-Cygni in \ion{C}{4} $\lambda 1549$; 
 
\item weak \ion{C}{4} $\lambda 1549$ in P-Cygni;  
 
\item absence of P-Cygni features.  
 
\end{enumerate} 
 
In Fig. \ref{group1} we show stars that are representatives of the first 
group. The total number of objects is 11: NGC 6629, NGC 6891, Hen 2-12,  
PB 8, Hen 2-108, NGC 6543, NGC 6567, NGC 6572, Vy 2-1, Hb7, and IC 4776.  
As can be seen, the \ion{C}{4} $\lambda 1549$ transition  
is often accompanied by \ion{N}{5} $\lambda 1238$,  
also in P-Cygni. Some other lines present are \ion{He}{2} $\lambda 1640$  
(e.g. PB 8), \ion{Si}{4} $\lambda \lambda 1394, 1403$ (e.g. Hen 2-108),  
and \ion{C}{3} $\lambda 1909$ (e.g. NGC 6891). A feature near $\sim 1720$\AA\, 
is also exhibited by several objects, and it might be due to \ion{N}{4}  
or \ion{Si}{4}. Line identifications, measurements of equivalent widths  
and fluxes for all objects will be presented in the next section.  
 
Only two stars were assigned to the second group, namely,  
CN 3-1 and Hen 2-131. Their spectra are shown  
in Fig. \ref{group23} (top and 2nd panel). The \ion{C}{4}  
$\lambda 1549$ line is clearly weaker than in group 1.  
As will be described later in Section \ref{vinfsec}, these two stars present also  
significantly lower terminal velocities. Despite weak,  
the transitions \ion{Si}{4} $\lambda \lambda 1394, 1403$ can also  
be identified. The remaining objects are: J900, IC 4997, NGC 5873, NGC 6818, NGC 
6803, and M 1-71. They compose the third group, and some of them are also shown  
in Fig. \ref{group23} (3rd to bottom panel). It can be readily noted the  
presence of strong emissions and of a weak and flat continuum. No P-Cygni  
emissions are seen. These characteristics suggest that  
only nebular features were detected by the {\it IUE} observations.  
Other archives than the ones listed in Table \ref{objects} were  
examined, but they all present similar spectra. 
An exception is the case of IC 4997. By analyzing another {\it IUE} 
archival spectrum (SWP08578) besides the one shown in Fig. \ref{group23} (SWP31683), we  
could see a slightly higher continuum and a \ion{N}{5} $\lambda 1238$ line 
possibly in P-Cygni. We chose to keep this star in group 3 since the
\ion{C}{4} $\lambda 1549$ line is not in P-Cygni in any other 
spectrum, including the ones in the high resolution mode (e.g. SWP41947). 
We have also included the star M 1-71 in group 3, although it shows only 
a featureless UV spectrum. For this object however, there is only one {\it IUE} 
file available, and it is possible that the central star was not 
within the {\it IUE} aperture during the observation. 

One object does not fit in our division scheme, namely, IC 5217.  
The \ion{C}{4} $\lambda 1549$ line in this star is intense, but not in P-Cygni. 
On the other side, a P-Cygni is found in \ion{N}{5} $\lambda 1238$, and also in
\ion{O}{5} $\lambda 1371$, although somewhat weak. Furthermore, this star does 
not present a flat continuum. We conclude that the wind spectrum is clearly
visible in this star, but it is highly contaminated by nebular emissions. 
 
We should emphasize that the three groups above were  
introduced by analyzing the $\sim 1150-2000$\AA\, interval, 
and that although we have shown in Figs. \ref{group1} and 
\ref{group23} low resolution spectra, the characteristics 
of these groups are confirmed from the available high dispersion data.
We chose the $\sim 1150-2000$\AA\, region for two reasons. First, it is in this wavelength range  
that we have found practically all the P-Cygni lines. Obviously, such 
profiles are formed in the stellar wind of the central star - which is our  
main interest - and not in the nebula. Second, the $\sim 2000-3000$\AA\,  
interval sometimes have no or very few features (see next section), whose  
nature (nebular or stellar) we cannot decide from low resolution data. 
The exception is the third group, where we have found various transitions, 
but as we previously argued, they are most likely nebular.
 
In addition to Figs. \ref{group1} and \ref{group23}, we show in Fig.  
\ref{wcpg} the {\it IUE} spectra of the two proto-type  
[WC]-1159 stars: A30 and A78, also at the same resolution.  
It is clear that these two objects have a spectrum  
that is remarkably different from the WELS  
spectra. The three P-Cygni lines \ion{N}{5} $\lambda 1238$,  
\ion{O}{5} $\lambda 1371$, and \ion{C}{4} $\lambda 1549$ are  
simultaneously present and more intense than in the WELS.  
Furthermore, the absorption part of these P-Cygni profiles  
(mainly in \ion{C}{4} $\lambda 1549$) are considerably broader,  
suggesting large terminal wind velocities (see Section \ref{vinfsec}).   
Although the E(B-V) for these stars have a role in  
determining the slope of the continuum, it is also  
evident that the flux is higher in the blue  
($\sim 1150$\AA) than in the red part of the {\it IUE}  
spectrum ($\sim 2000$\AA). This is not seen in most of the  
WELS, and might indicate that [WC]-PG 1159 stars have higher  
effective temperatures. 
 
We have assigned the stars NGC 6567, NGC 6572, and NGC 6543,  
to the group presenting intense \ion{C}{4} $\lambda 1549$ in  
P-Cygni. However, in addition to the \ion{C}{4} $\lambda 1549$ line  
exhibited by the stars in group 1, these three objects also show 
simultaneously \ion{N}{5} $\lambda 1238$ and \ion{O}{5} $\lambda 1371$ P-Cygni 
emissions. This characteristic resembles the [WC]-PG 1159 class and is shown 
in Fig. \ref{wcpg2}. Thus, although a comparison between spectra shown 
in Fig. \ref{group1}, \ref{group23}, and \ref{wcpg} argue against 
those WELS being [WC]-PG 1159 stars, we cannot  
be sure regarding the objects shown in Fig. \ref{wcpg2}. 
In fact, the central star of NGC 6567 for example, was already 
considered as a [WC]-PG 1159 object by Hamann (1996). 
This point will be addressed later in the paper. 
 
\subsection{Line Identifications and Measurements} 
 
In Tables 2, 3, 4 and 5 we present line identifications, measurements  
of equivalent widths and line fluxes of all {\it IUE} low resolution spectra  
considered in this work. For completeness, we have included IC 5217 in Table
4, although this object does not belong to the group of WELS without P-Cygni 
emissions (group 3). The equivalent widths were measured by the conventional  
method of adjusting a Gaussian function to the line profile. When this  
was not possible, we have computed the line area above the adopted  
continuum. In the case of a P-Cygni profile, only its emission feature  
was considered in the measurement. We emphasize that our intention is  
not to construct an UV atlas for the WELS. Therefore, we have only  
listed and measured the most conspicuous lines in their spectra.  
It should also be kept in mind that throughout this paper we make  
reference to lines such as \ion{C}{4} $\lambda 1549$ and  
\ion{N}{5} $\lambda 1238$, but the exact atomic transitions  
might be slightly different. In these two cases for example, the carbon and  
nitrogen lines are actually the doublets: \ion{C}{4} $\lambda \lambda 1548, 51$  
and \ion{N}{5} $\lambda \lambda 1238, 42$. 
 
Previous studies have investigated the UV spectrum of some stars seen in Tables 2-5.  
In general, we have found a good agreement between our results and the  
ones that also present line measurements. For Vy 2-1, NGC 6629, M 1-71, and Hen 2-108, we found  
no reference in the literature regarding their {\it IUE} spectra from the MAST  
database. Although we can find references for the stars CN 3-1, NGC 6891,  
J900, Hen 2-12, NGC 5873, and NGC 6803, they generally involve  
a large sample of objects and are not focused on line identifications,  
measurements or the WELS evolutionary state. Details concerning  
some individual objects are discussed below.  
 
{\bf NGC 6567:} Although there are four references regarding this object  
in the MAST database, only the one of Hyung et al. (1993) have focused  
in some detail the main features of its ultraviolet spectrum.  
For \ion{C}{4} $\lambda 1549$, these authors found a line flux (in  
units of $10^{-13}$ ergs s$^{-1}$ cm$^{-2}$) of 12.8, while we  
have measured 13 $\pm$ 1. Taking into consideration the uncertainty  
in the line measurements of their work (of $\sim 15\%$), our results  
present a very reasonable agreement for this line and others such  
as \ion{N}{5} $\lambda 1238$, \ion{O}{5} $\lambda 1371$,  
\ion{C}{2} $\lambda 2325$, and \ion{Mg}{2} $\lambda 2798$.   
 
{\bf NGC 6572:} There are several references to the UV spectrum of this 
object. The most important to our discussion is the one of Hyung et al.  
(1994), which presents a detailed high resolution spectral study from  
the UV to the optical, with several line identifications and measurements.  
Their main aim was however to explore diagnostic lines in the rich  
spectrum of NGC 6572 in order to determine plasma parameters, and  
not to discuss its evolutionary status.  
 
{\bf NGC 6543:} The {\it IUE} spectrum of this star was studied before  
by Bianchi et al. (1986) and Perinotto et al. (1989). Both works  
have heavily focused only in the determination of stellar and wind  
parameters (e.g. $T_{eff}$ and $\dot{M}$). No comparison to other  
objects as in the context of the present paper was made. 
 
{\bf PB 8:} This planetary nebula was studied together with other  
7 objects by Feibelman (2000). Line fluxes provided in his work  
are in good agreement with our measurements. For \ion{N}{5}  
$\lambda 1238$ for example, we encountered $\sim 2.6$ (units of  
10$^{-13}$ ergs cm$^{-2}$ s$^{-1}$) while Feibelman derived  
2.4. Although the emission in $\sim 1724$\AA\, was identified  
in his work as \ion{N}{4} $\lambda 1719$, it is possible that  
it arises actually from \ion{Si}{4} $\lambda 1722$. In fact, we 
have also identified two lines in 1394\AA\, and 1403\AA\, which 
are likely to stem from \ion{Si}{4}.
 
{\bf Hen 2-131:} Its {\it IUE} spectra were briefly discussed  
by Adams \& Seaton (1982). However, the attention was given to the 
determination of the C/O ratio in the nebula, and not to the central 
star. 
 
{\bf IC 4776:} This object was studied by Herald \& Bianchi  
(2004) by means of expanding atmosphere models, and its  
physical parameters and chemical abundances were obtained.  
As far as we are concerned, their work constitute the first  
to analyze the stellar wind of a WELS with a state-of-art  
non LTE code (CMFGEN; Hillier \& Miller 1998). The same main  
observed wind emissions were identified in their and our  
paper. The \ion{O}{5} $\lambda 1371$ is predicted by their model, but  
it is very weak or absent in the {\it IUE} spectrum. Herald \& Bianchi  
(2004) have also analyzed the star A78, and a [WC]-PG 1159 classification  
was confirmed. Surprisingly, the physical parameters obtained for A78  
and IC 4776 are quite different. A78 is almost twice hotter ($T_{eff} \sim 113kK$)  
and present a larger terminal velocity (3200 km s$^{-1}$) than IC 4776. 
These findings support the idea that these two stars do not belong  
to the same spectroscopy class, i.e., that WELS are not [WC]-PG 1159 stars  
(see Section \ref{discussion}). 
 
{\bf Hb7:} Gauba et al. (2001) also analyzed the low {\it IUE} resolution  
spectrum of this star, together with another planetary nebula (Sp 3). Some  
wind parameters (e.g. $\dot{M}$) as well as the effective temperature and  
core mass were estimated, but no atmosphere models were used. The spectral  
energy distribution from the far infrared to the ultraviolet was also  
investigated. No line identifications and measurementes were presented  
and the status of Hb7 as a WELS was not discussed. 
 
{\bf NGC 6818:} The spectrum of this object (nebular plus central star)  
was extensively studied by Hyung et al. (1999). We refer the reader to  
their work for a more detailed analyses than the one presented here.  
The chemical composition of the planetary nebula was further explored  
by the work of Pottasch et al. (2005). 
 
{\bf IC 5217:} A very similar analysis to NGC 6818 (Hyung et al. 1999)  
was done in the case of IC 5217 by Hyung et al. (2001). Again, we refer the reader  
to their work for detailed line identifications and measurements. We  
highlight that in both references the identity of these two stars  
as WELS is not discussed, and emphasis is given mainly to the  
determination of plasma parameters such as the electronic  
temperature, density, ionic concentrations, and chemical  
abundances. 
 
\section{TERMINAL VELOCITIES}   
\label{vinfsec} 
 
In this section, we derive terminal velocities ($v_\infty$) for all 
WELS presenting P-Cygni profiles in \ion{N}{5} $\lambda 1238$ and/or \ion{C}{4} $\lambda
1549$. The $v_\infty$, together with the mass-loss rate ($\dot{M}$), constitute 
the most important physical parameters describing the stellar winds of hot
stars. Their empirical determination are often used to test the theory of 
radiatively driven stellar winds, to compare stars of different spectral 
classes, and also as an input in stellar evolution models and interstellar 
medium studies (see for example Prinja et al. 1990; Lamers \& Cassinelli 
1999). The determination of $\dot{M}$ for the WELS presenting wind features 
is beyond the scope of the present paper, since it requires sophisticated 
non LTE expanding atmosphere models for reliable estimates. Work is under 
way by our group to accomplish this goal.

Because only a few high resolution spectra are available, most of the 
terminal velocities of the WELS can only be obtained from low 
dispersion {\it IUE} spectra. A priori, an accurate  
determination of $v_\infty$ from a $\sim 6$\AA\, resolution spectrum  
is a complicated task, since we generally do not see an extended  
and saturated absorption in a P-Cygni line. Nevertheless, in order to  
overcome this difficulty, Prinja (1994) provided a relation between  
$\Delta \lambda$ (defined below, obtained at low resolution)   
and $v_\infty$ (obtained at high resolution) by analyzing a large  
sample of different hot stars (O, B, WR, and also CSPN). The resultant  
calibration of his work is represented by:  
 
\[ v_\infty = a_1 + a_2[\Delta \lambda] + a_3[\Delta \lambda]^2.   \] 
 
In this equation, $\Delta \lambda$ ($= \lambda _{peak} - \lambda _{min}$)  
is the wavelength difference between the peak and the absorption minima  
of a low resolution P-Cygni profile, and $a_1$, $a_2$ and $a_3$ have  
different values for \ion{C}{4} $\lambda 1549$ and \ion{N}{5} $\lambda 1238$  
(see Prinja 1994). This calibration is perfectly suitable for our purpose,  
and obeys the definition of the terminal velocity from the black absorption  
core ($v_{black}$) of a P-Cygni line (Prinja et al. 1990). 
 
In Table 6 we present the terminal velocities derived for the WELS, 
and also for the [WC]-PG 1159 stars A30 and A78. We used the usual method for 
WELS with high resolution spectra  (using the Doppler effect and $v_{black}$) 
and the equation described above for low resolution spectra. We estimate 
the errors in the determination of $v_\infty$ to be $\sim 20\%$. 
 
The most important thing to be noted in Table 6 is that A30 and A78  
have much higher terminal velocities than the WELS. While the majority of  
WELS have values concentrated in the $\sim 1000-1500$ km s$^{-1}$ interval,  
these two [WC]-PG 1159 stars have a terminal velocity of $\sim 3000$ km  
s$^{-1}$ (from \ion{C}{4}). Such high differences cannot be explained 
by the uncertainties in the measurements. Although different  
works in the literature have estimated $v_\infty$ for A30 and A78, we  
have also measured this quantity in their low resolution spectra to follow  
the same method applied for the other stars (Prinja's calibration). Among  
the previous studies regarding A30, we highlight the work of Harrington \&  
Feibelman (1984), who obtained a terminal velocity of $\sim 4000$ km 
s$^{-1}$. At that time however, the $v_\infty$ determination was based on  
the largest negative velocity seen in a P-Cygni profile  
(when the line returns to the continuum). This procedure tends to obtain  
larger values than the $v_{black}$ method (Prinja et al. 1990). The central  
star of A78 was recently analyzed by sophisticated non LTE atmosphere models  
by Herald \& Bianchi (2004). These authors found that a velocity  
of $\sim 3200$ km s$^{-1}$ could satisfactorily represent $v_\infty$, a value  
slightly higher than ours. 
 
Other informations can be obtained by analyzing Table 6.  
It is clear for example, that CN 3-1 and Hen 2-131 have  
lower terminal velocities than most of the WELS. Indeed,  
as we have shown in Section \ref{uvwels} (Fig. \ref{group23}),  
their UV spectra are quite different. 

In order to further illustrate the differences between the terminal 
velocities of the WELS and the two prototype [WC]-PG 1159 stars A30 and A78, 
as shown in Table \ref{vinf}, and to compare the results obtained to other hydrogen deficient CSPN, we 
present in Fig. \ref{hist} the $v_\infty$ distribution for several 
stars belonging to the spectroscopic classes [WCL], [WCE], [WC]-PG 1159, 
WELS, and PG 1159. Once again, it can be seen that the terminal 
velocities measured for the WELS are mainly concentrated between $\sim 1000-1500$ km s$^{-1}$. 
Moreover, their $v_\infty$ tends to be higher and lower than in the 
[WCL] and [WCE] class, respectively. It is also clear that the [WC]-PG 1159 
class (represented by the two prototype stars A30 and A78) and the 
few PG 1159 stars that show a stellar wind, have the highest terminal 
velocities among the hydrogen deficient CSPN. 
 
%%%%%%%%%%%%%%%%%%%%%%%%%%%%%%%%%%%%%%%%%%%%%%%%%%%%%%%%%%%%%%%%%%%%%   
   
\section{DISCUSSION AND CONCLUSIONS} 
\label{discussion} 
   
From a comparison between several optical spectra of WELS, [WC]-PG 1159,  
and PG 1159 stars, Parthasarathy et al. (1998) have proposed that  
the WELS are actually [WC]-PG 1159 stars. This claim was not  
confirmed by further studies and as we mentioned,  
some authors have warned about this assertion until a more  
comprehensive study is achieved (Werner \& Herwig 2006). After an  
analysis of the main UV characteristics of the WELS and the two  
prototype [WC]-PG 1159 stars A30 and A78, our next step was  
to compare the results obtained for these two class of objects in  
order to address this important issue. 
 
\begin{figure}[!ht]   
\centering   
\epsscale{.9}
\plotone{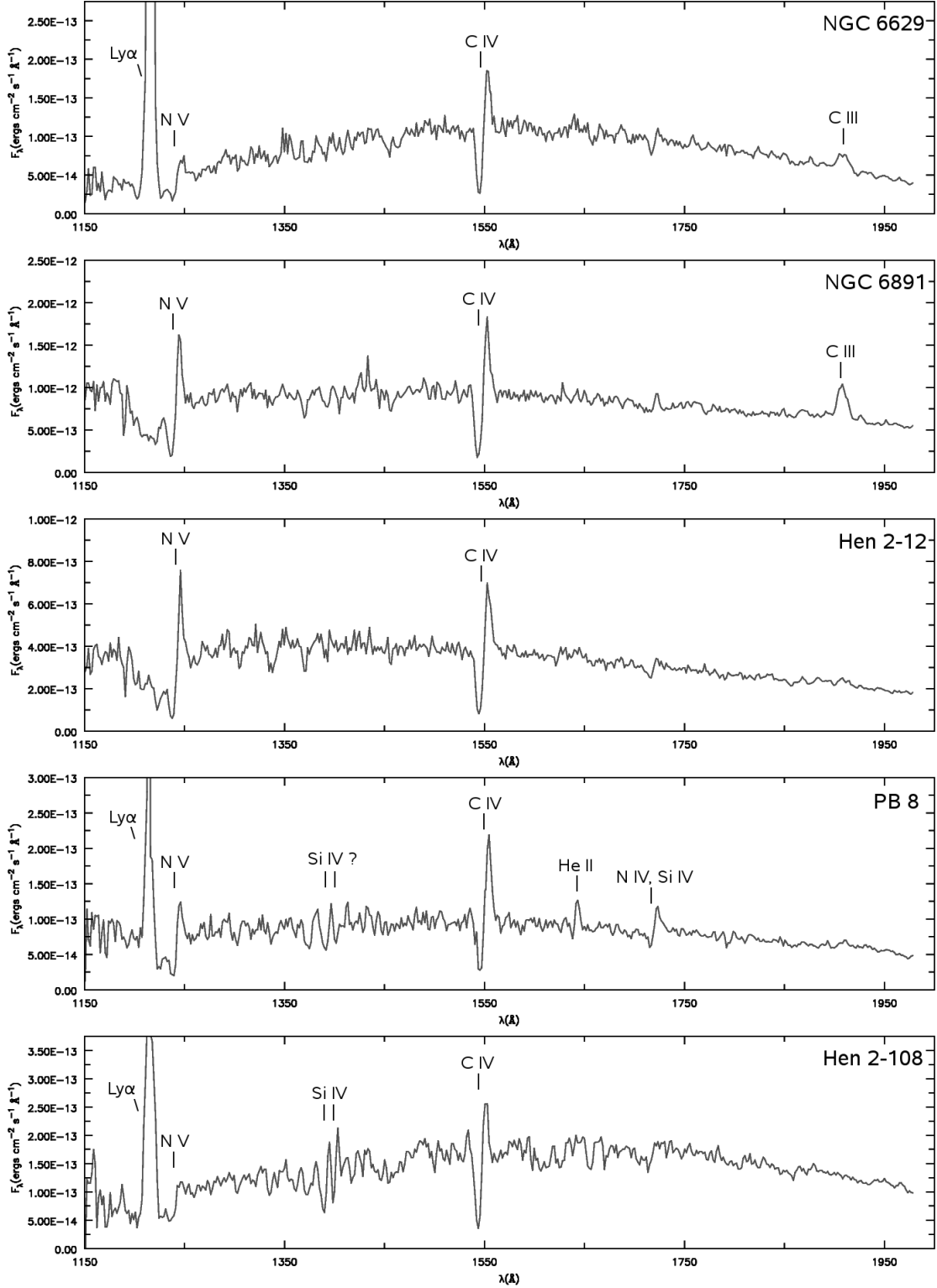}
\caption{WELS presenting strong P-Cygni profiles (group 1) in \ion{C}{4} 
  $\lambda 1549$. The \ion{N}{5} $\lambda 1238$ is often present and also in 
  P-Cygni. Low resolution {\it IUE} spectra ($\sim 6$\AA).}   
\label{group1}   
\end{figure}

\begin{figure}[!ht]   
\centering   
\plotone{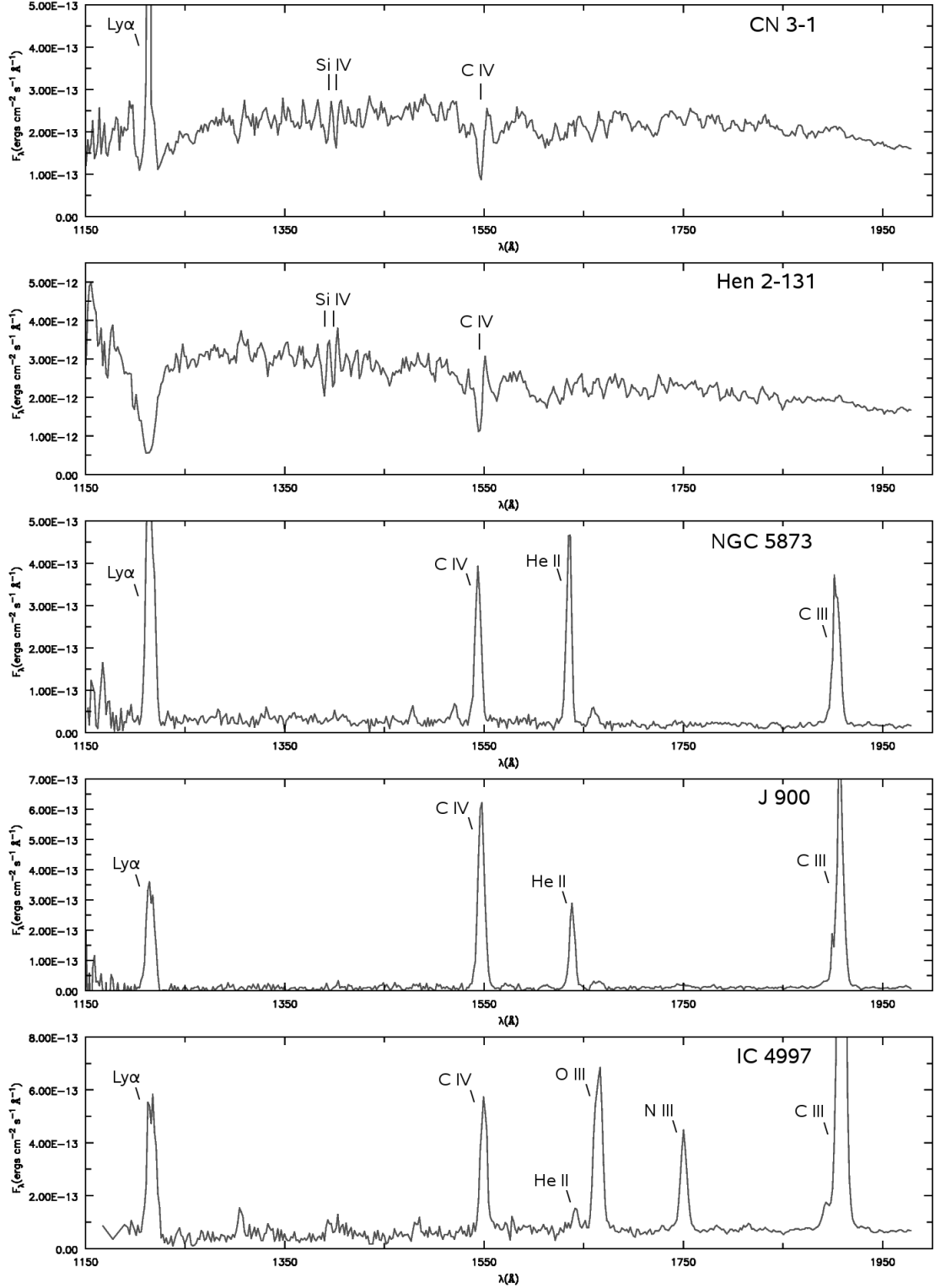}
\epsscale{1}
\caption{WELS presenting weak P-Cygni profiles in  \ion{C}{4} $\lambda 1549$  
(group 2): top and 2nd panels. WELS presenting absence of P-Cygnis   
  and very weak continuum (group 3): 3rd to 5th panels. Low resolution {\it IUE} spectra ($\sim 6$\AA).}   
\label{group23}   
\end{figure}

\begin{figure}[!ht]   
\centering   
\includegraphics[width=18cm, height=10cm]{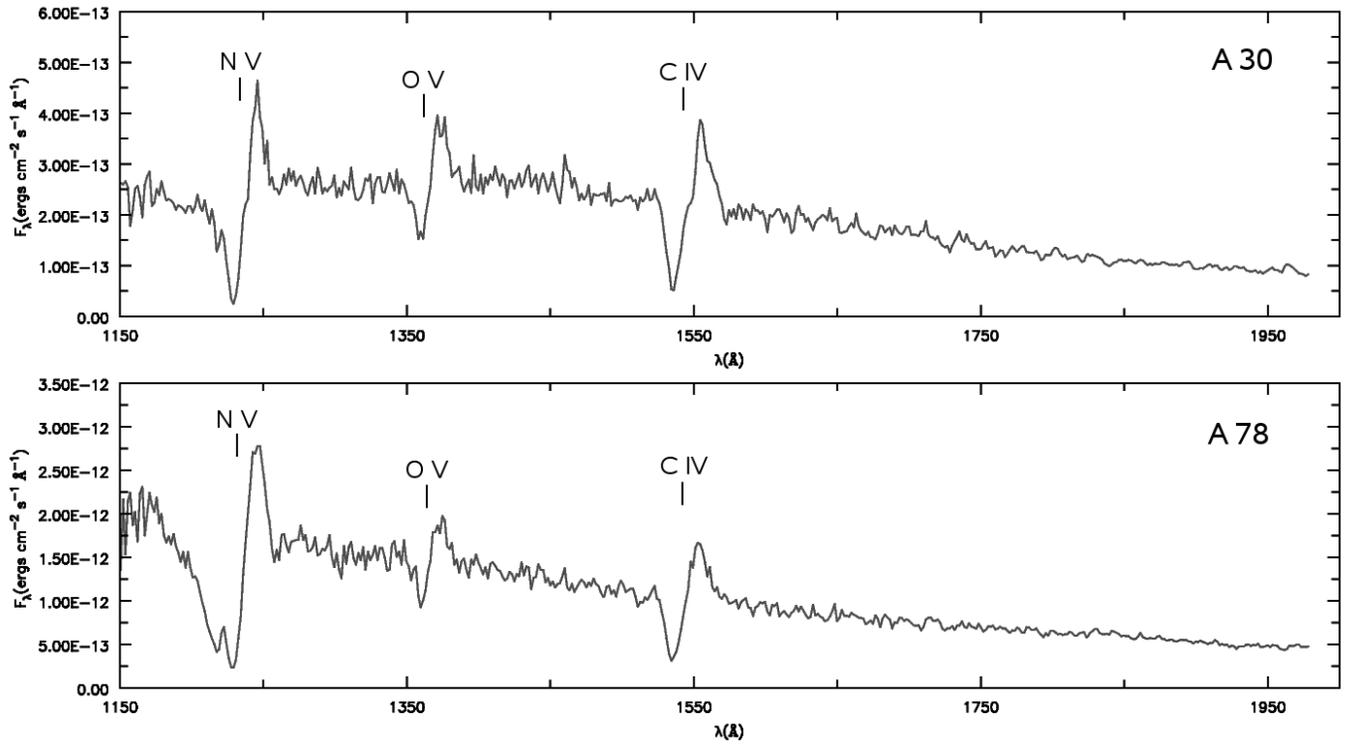}   
\caption{Low resolution {\it IUE} spectra of the two prototype [WC]-PG 1159 stars - A 30 (top) and A 78 (bottom).}   
\label{wcpg}   
\end{figure}

\begin{figure}[!ht]   
\centering   
\includegraphics[width=17cm, height=13cm]{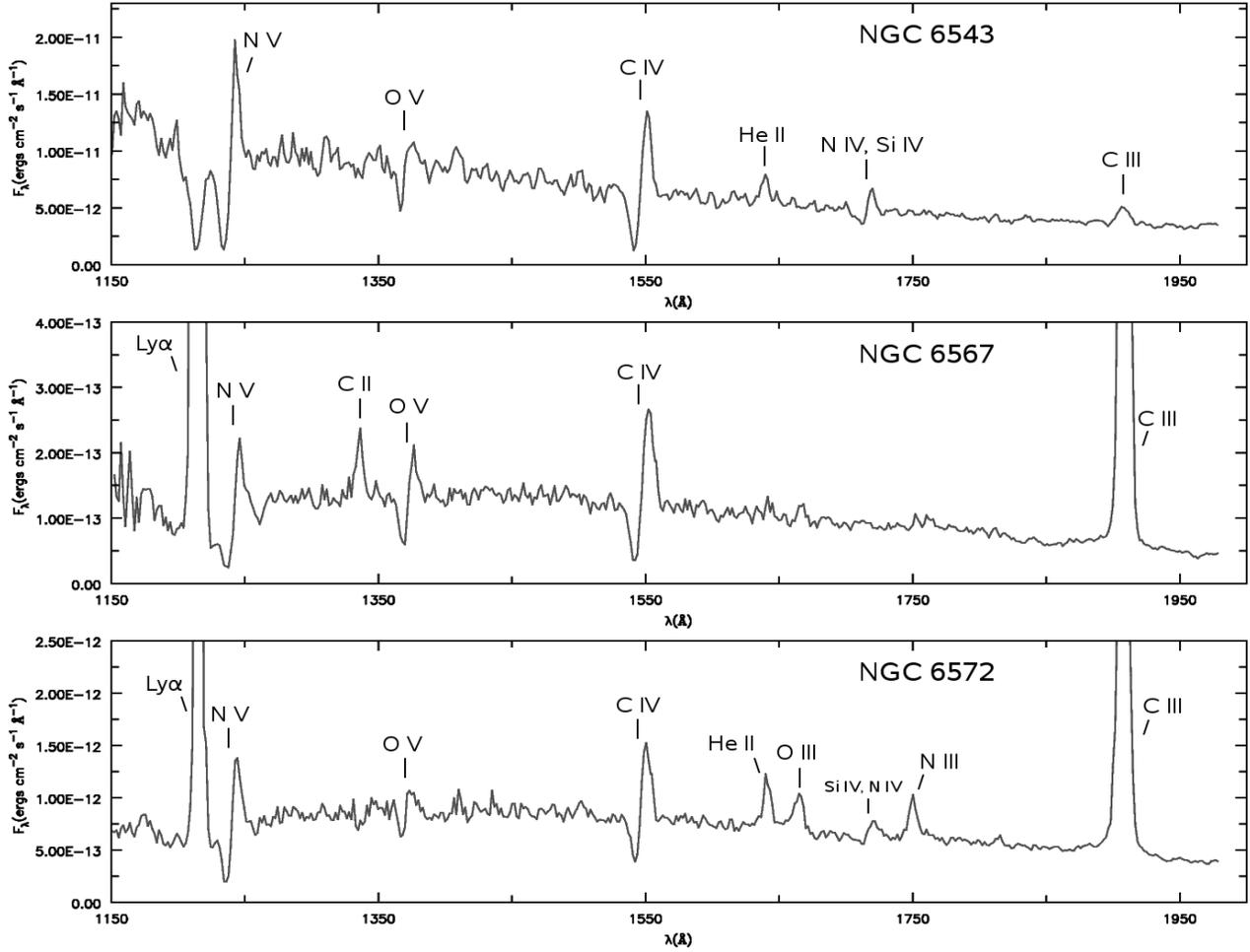}   
\caption{Low resolution {\it IUE} spectra of WELS resembling [WC]-PG 1159  
stars: NGC 6543, NGC 6567, and NGC 6572.}   
\label{wcpg2}   
\end{figure}

\begin{figure}[!ht]   
\centering   
\includegraphics[width=15cm, height=15cm]{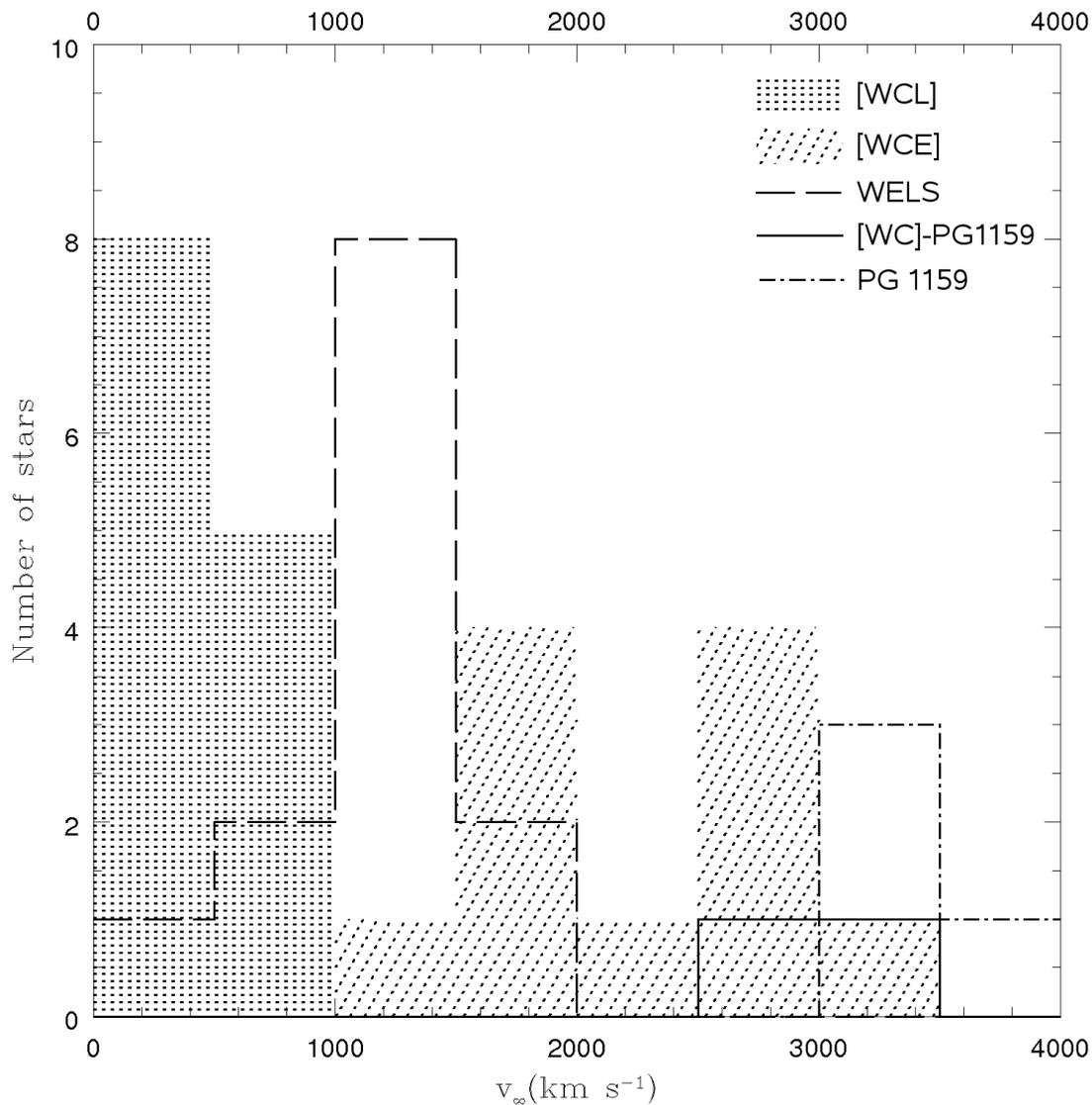}   
\caption{Distribution of terminal velocities of [WCL], [WCE], WELS, 
[WC]-PG 1159, and PG 1159 stars. The data for the WELS and 
[WC]-PG 1159 stars were homogeneously obtained from low resolution 
{\it IUE} spectra using \ion{C}{4} $\lambda 1549$ and the calibration 
of Prinja (1994). The terminal velocities for the [WCL], [WCE] and 
PG 1159 stars are from Koesterke (2001), and were obtained by means of non 
LTE expanding atmosphere models. The bin width is 500 km s$^{-1}$.}   
\label{hist}   
\end{figure}

As we have shown, most of the WELS present a UV spectrum considerably  
different from the [WC]-PG 1159 stars. While this last class present simultaneously  
P-Cygni profiles in \ion{N}{5} $\lambda 1238$, \ion{O}{5} $\lambda 1371$, and  
\ion{C}{4} $\lambda 1549$, the majority of WELS present a very weak or no  
\ion{O}{5} $\lambda 1371$ (see Fig. \ref{group1}). The same is true at least for  
some objects regarding the \ion{N}{5} $\lambda 1238$ line. 
The only exceptions are the objects NGC 6543, NGC 6567, and NGC 6572.  
Their spectra in fact resemble the ones of the [WC]-PG 1159 stars  
(see Figs. \ref{wcpg} and \ref{wcpg2}): their \ion{O}{5} $\lambda 1371$  
line is clearly visible as well as the other transitions mentioned.  
 
Besides the spectral differences found in the ultraviolet part of  
the spectrum, we have also found that the terminal velocities of the  
WELS are considerably lower than in [WC]-PG 1159 stars. Our  
Table 6 shows that the bulk of the WELS have $v_\infty$  
between $\sim 1000-1500$ km s$^{-1}$, and A30 and A78 have  
values about $3000$ km s$^{-1}$. This difference might  
represent different physical parameters underlying these two  
class of stars. The theory of radiatively driven stellar winds  
for example, predicts that the terminal velocity of a star is  
related with the escape velocity ($v_{esc}$), which in turn  
depend on other physical parameters (e.g. mass and radius; Abbott 1978; Lamers  
\& Cassinelli 1999). Moreover, $v_\infty$ is also known to correlate with the  
effective temperature ($T_{eff}$) in several class of stars  
(see Fig. 8 of Prinja et al. 1990). The $T_{eff}$ tends  
to be higher for stars with high terminal velocities. 
 
From the considerations above described, we conclude that the  
[WC]-PG 1159 stars are distinct from the WELS, in contrast  
with the claim made by Parthasarathy et al. (1998) on the  
basis of optical spectroscopy. It should be noted however,  
that the situation for the central stars NGC 6543, NGC 6567 and  
NGC 6572 is ambiguous. From one side, they have a spectrum  
compatible with the [WC]-PG 1159 class. On the other hand, they do  
not present high terminal velocities as it is the case of  
A30 and A78. From both low and high resolution data it is  
obtained $v_\infty$ values less than $2000$ km s$^{-1}$ for  
these three stars (see Table 6).  
 
If the WELS are not [WC]-PG 1159 stars, what is their 
role in the evolutionary sequence [WR] $\rightarrow$ PG 1159 ?  
Do they form an alternative channel of evolution ? In order  
to elucidate these and other similar questions, and to further  
clarify the differences between them and the [WC]-PG 1159 stars,  
we clearly need to determine their physical parameters and  
chemical abundances. Non LTE expanding atmosphere models are  
being computed with this purpose by our group with  
the CMFGEN code of Hillier \& Miller (1998).  
In this way, their position in the HR diagram could be  
determined, a more efficient comparison to [WR], [WC]-PG 1159  
and PG 1159 stars could be made, and their evolutionary status  
could be better determined. 
 
%%%%%%%%%%%%%%%%%%%%%%%%%%%%%%%%%%%%%%%%%%%%%%%%%%%%%%%%%%%%%%%   
 
%%%%%%%%%%%%%%%%%%%%%%%%%%%%%%%%%%%%%%%%%%%%%%%%%%%%%%%%%%%%%%%   
   
%%%%%%%%%%%%%%%%%%%%%%%%%%%%%%%%%%%%%%%%%%%%%%%%%%%%%%%%%%%%%%%   
   
%%%%%%%%%%%%%%%%%%%%%%%%%%%%%%%%%%%%%%%%%%%%%%%%%%%%%%%%%%%%%%%   
     
\begin{acknowledgements}   
   
W. M. acknowledges CNPq for financial support (151635-2005-6).    
H. J. acknowledges CNPq for financial support. 
   
\end{acknowledgements}   
   
{}   
   
%%%%%%%%%%%%%%%%%%%%%%%%%%%%%%%%%%%%%%%%%%%%%%%%%%%%%%%%%%%%%%%%%   

\clearpage
\begin{deluxetable}{lllcc}
\tabletypesize{\small}
\tablecaption{Observational Data Utilized.}
\tablewidth{0pt}
\tablehead{
\colhead{PN G} & \colhead{Star} & \colhead{Data Set} &
\colhead{Resolution} & \colhead{Aperture}
}

\startdata
002.0-13.4 & IC 4776   &  SWP16504,LWP13842 &  Low &  Large \\   
003.9-14.9 & HB 7      &  SWP52257 &  Low &  Large \\   
007.0-06.8 & Vy2-1     &  SWP44200 &  Low &  Large \\   
009.4-05.0 & NGC 6629  &  SWP35968,LWP15329 &  Low &  Large \\   
011.7-00.6 & NGC 6567  &  SWP45362,LWP23708 &  Low &  Large \\   
025.8-17.9 & NGC 6818  &  SWP01704,LWR10557 &  Low &  Large \\   
034.6+11.8 & NGC 6572  &  SWP42043,LWP20787,SWP42059 &  Low/High &  Large \\   
038.2+12.0 & CN 3-1    &  SWP31600,LWP11819 &  Low &  Large \\   
046.4-04.1 & NGC 6803  &  SWP06256,LWR05428 &  Low &  Large \\   
054.1-12.1 & NGC 6891  &  SWP08173,LWP23363,SWP33503 &  Low/High &  Large \\   
055.5-00.5 & M 1-71    &  SWP31678 &  Low &  Large \\   
058.3-10.9 & IC 4997   &  SWP31683,LWP20706 &  Low &  Large \\   
096.4+29.9 & NGC 6543  &  SWP54891,LWP24732,SWP03323 &  Low/High &  Large \\   
100.6-05.4 & IC 5217   &  SWP07257,LWR01785 &  Low &  Large \\   
194.2+02.5 & J 900     &  SWP53870,LWP29940 &  Low &  Large \\   
253.9+05.7 & HEN 2-12  &  SWP16346,LWR12565 &  Low &  Large \\   
292.4+04.1 & PB 8      &  SWP30476 &  Low &  Large \\   
315.1-13.0 & HEN 2-131 &  SWP47003,LWR05736,SWP07653 &  Low/High &  Large \\   
316.1+08.4 & HEN 2-108 &  SWP47513,LWP25380 &  Low &  Large \\   
331.3+16.8 & NGC 5873  &  SWP30150 &  Low &  Large \\   
\\
208.5+33.2 & A30       &  SWP07955 &  Low &  Small \\   
081.2-14.9 & A78       &  SWP44942,SWP19906 &  Low/High &  Large \\   
\\
\enddata   
\label{objects}
\end{deluxetable}

%%%%%%%%%%%%%%%%%%%%%%%%%%%%%%%%%%%%%%%%%%%%%%%%%%%%%%%%%%%%%%%%%%%%%%%%%%%%%%%%%%%%%%%%   

\clearpage

\begin{deluxetable}{lllcc}
\tabletypesize{\small}
\tablecaption{WELS with strong \ion{C}{4} $\lambda 1549$ in P-Cygni.}
\tablewidth{0pt}
\tablehead{
\colhead{{\bf Star}} & \colhead{$\lambda_{obs}$} & \colhead{  Transition } &
\colhead{ W$_{\lambda}$ } & \colhead{ Flux ($10^{-13}$ ergs s$^{-1}$ cm$^{-2}$) } 
}

\startdata

{\bf Vy 2-1} & 1246.3* & \ion{N}{5} $\lambda 1238$ & -13 $\pm$ 3    & 0.62 $\pm$ 0.08 \\   
             & 1552.9* & \ion{C}{4} $\lambda 1549$ & -6.4 $\pm$ 1.6 & 0.56 $\pm$ 0.08 \\    
             & 1721.0  & \ion{N}{4} $\lambda 1719$?, \ion{Si}{4} $\lambda 1722$? & -4.7 $\pm$ 1.1 & 0.42 $\pm$ 0.07 \\    
             & 1906.7  & \ion{C}{3} $\lambda 1909$ & -6.8 $\pm$ 1.2 & 0.5  $\pm$ 0.07 \\    
\\     
{\bf NGC 6543} & 1243.6* & \ion{N}{5} $\lambda 1238$ & -6.0 $\pm$ 0.8 &  540 $\pm$ 40 \\   
               & 1376.4* & \ion{O}{5} $\lambda 1371$ & -3 $\pm$ 1     &  300 $\pm$ 100 \\
               & 1551.6* & \ion{C}{4} $\lambda 1549$ & -8 $\pm$ 1     &  510 $\pm$ 50 \\   
               & 1639.8  & \ion{He}{2} $\lambda 1640$& -3.7 $\pm$ 0.4 &  200 $\pm$ 20 \\ 
               & 1720.0* & \ion{N}{4} $\lambda 1719$?, \ion{Si}{4} $\lambda 1722$?    
               &  -2.5 $\pm$ 0.3 &  110 $\pm$ 10 \\   
               &  1907.8 & \ion{C}{3} $\lambda \lambda 1909$ & -4.9 $\pm$ 0.8 & 170 $\pm$ 20 \\
\\ 
{\bf NGC 6567} & 1246.5* & \ion{N}{5} $\lambda 1238$ & -6 $\pm$ 1 & 7 $\pm$ 1 \\   
               &  1336.1 & \ion{C}{2} $\lambda 1334-36$ & -4.3 $\pm$ 0.9 & 5.8 $\pm$ 0.9 \\   
               & 1377.2* & \ion{O}{5} $\lambda 1371$ & -3.6 $\pm$ 0.3 & 4.5 $\pm$ 0.3 \\  
               & 1553.4* & \ion{C}{4} $\lambda 1549$ & -11 $\pm$ 1 & 13 $\pm$ 1 \\  
               &  1667.1 & \ion{O}{3} $\lambda 1666-69$ & -1.2 $\pm$ 0.5 & 1.2 $\pm$ 0.4 \\      
               & 1908.7  & \ion{C}{3} $\lambda 1909$ & saturated &  saturated \\  
               & 2326.7  & \ion{C}{2} $\lambda 2325$ & -27 $\pm$ 4 & 6.3 $\pm$ 0.4 \\   
               & 2798.4  & \ion{Mg}{2} $\lambda 2796-98$ & -15 $\pm$ 1 & 6.6 $\pm$ 0.4 \\   
\\
{\bf NGC 6572} & 1244.6* & \ion{N}{5} $\lambda 1238$ & -7.1 $\pm$ 0.9 & 47 $\pm$ 4 \\   
               & 1375.4* & \ion{O}{5} $\lambda 1371$ & -1.8 $\pm$ 0.4 & 16 $\pm$ 3 \\   
               & 1551.5* & \ion{C}{4} $\lambda 1549$ & -6.4 $\pm$ 0.7 & 51 $\pm$ 4 \\   
               & 1641.3  & \ion{He}{2} $\lambda 1640$ & -4.5 $\pm$ 0.6 & 33 $\pm$ 3 \\   
               & 1664.6  & \ion{O}{3} $\lambda 1666-69$ & -4.2 $\pm$ 0.4 & 29 $\pm$ 2 \\  
               &  1721.5 & \ion{N}{4} $\lambda 1719$?, \ion{Si}{4} $\lambda 1722$? & -1.7 $\pm$ 0.4 &  10 $\pm$ 2 \\   
               &  1750.7 & \ion{N}{3} $\lambda 1751$ & -4.7 $\pm$ 0.4 & 29 $\pm$ 2 \\   
               &  1815.3 & \ion{Ne}{3} $\lambda 1815$& -0.8 $\pm$ 0.1 & 4.4 $\pm$ 0.5 \\   
               &  1907.9 & \ion{C}{3} $\lambda 1909$& saturated & saturated \\   
               &  2169.0 & \ion{C}{3} $\lambda 2163$? & -12 $\pm$ 8 &  21 $\pm$ 7 \\   
               &  2293.6 & \ion{C}{3} $\lambda 2293-97$ & -3 $\pm$ 1 & 7 $\pm$ 2 \\   
               &  2326.9 & \ion{C}{2} $\lambda 2325$ & -46 $\pm$ 5 & 117 $\pm$ 10 \\ 
               &  2470.3 & \ion{O}{2} $\lambda 2470$ & -18 $\pm$ 1 &  59 $\pm$ 2 \\   
\\
{\bf NGC 6891} & 1245.3* & \ion{N}{5} $\lambda 1238$ & -5.8 $\pm$ 0.7 & 45 $\pm$ 3 \\   
               & 1434.0  & \ion{C}{1} $\lambda 1432$?& -1.4 $\pm$ 0.7 & 13 $\pm$ 5 \\   
               & 1552.9* & \ion{C}{4} $\lambda 1549$ & -5.1 $\pm$ 0.8 & 47 $\pm$ 5 \\ 
               & 1723.0  & \ion{N}{4} $\lambda 1719$?, \ion{Si}{4} $\lambda 1722$? & -1.3 $\pm$ 0.2 &  9 $\pm$ 1 \\    
               & 1908.0 & \ion{C}{3} $\lambda 1909$  & -6.8 $\pm$ 0.5 & 44 $\pm$ 2 \\    
\\   
{\bf NGC 6629} &  1247.8* & \ion{N}{5} $\lambda 1238$ & -5.5 $\pm$ 2.4 & 2.3 $\pm$ 0.8 \\   
               &  1554.0* & \ion{C}{4} $\lambda 1549$ & -4.4 $\pm$ 0.3 & 4.6 $\pm$ 0.2 \\
               &  1724.3* & \ion{N}{4} $\lambda 1719$?, \ion{Si}{4} $\lambda 1722$? & -0.9 $\pm$ 0.2 &  0.8 $\pm$ 0.2 \\
               &  1908.5  & \ion{C}{3} $\lambda 1909$ & -4.3 $\pm$ 0.6 & 2.5 $\pm$ 0.3 \\  
\\   
{\bf Hen 2-12} & 1246.4* & \ion{N}{5} $\lambda 1238$ & -5.2 $\pm$ 0.8 & 17 $\pm$ 2 \\ 
               & 1554.2* & \ion{C}{4} $\lambda 1549$ & -5.5 $\pm$ 0.4 &  20 $\pm$ 1 \\  
               & 1723.4* & \ion{N}{4} $\lambda 1719$?, \ion{Si}{4} $\lambda 1722$? & -0.7 $\pm$ 0.2 & 2.1 $\pm$ 0.7 \\  
\\  
{\bf Pb 8}     & 1245.9* & \ion{N}{5} $\lambda 1238$ & -3.6 $\pm$ 0.5 & 2.6 $\pm$ 0.3 \\  
               & 1382.2* & \ion{O}{5} $\lambda 1371$?& -2.2 $\pm$ 0.4 & 1.8 $\pm$ 0.3 \\
               & 1397.1* & \ion{Si}{4} $\lambda 1394$? & -1.2 $\pm$ 0.1 & 1.0 $\pm$ 0.1 \\      
               & 1413.1* & \ion{Si}{4} $\lambda 1403$? & -1.3 $\pm$ 0.3 &  1.1 $\pm$ 0.2 \\   
               & 1554.5* & \ion{C}{4} $\lambda 1549$ & -7.5 $\pm$ 0.5 & 7.1 $\pm$ 0.3 \\ 
               & 1643.4* & \ion{He}{2} $\lambda 1640$ & -2.1 $\pm$ 0.2 & 1.9 $\pm$ 0.1 \\
               & 1724.1* & \ion{N}{4} $\lambda 1719$?, \ion{Si}{4} $\lambda 1722$? & -2.7 $\pm$ 0.5 &  2.2 $\pm$ 0.4 \\
\\   
{\bf Hen 2-108} & 1245.1* & \ion{N}{5} $\lambda 1238$ & very weak or absent & very weak or absent \\   
                & 1394.9* & \ion{Si}{4} $\lambda 1394$ & -1.5 $\pm$ 0.5 & 1.8 $\pm$ 0.5 \\   
                & 1403.8* & \ion{Si}{4} $\lambda 1403$ & -1.7 $\pm$ 0.4 & 2.2 $\pm$ 0.4 \\
                & 1551.9* & \ion{C}{4} $\lambda 1549$  & -2.9 $\pm$ 0.4 & 4.6 $\pm$ 0.5 \\  
\\ 
{\bf IC 4776}   & 1245.3* & \ion{N}{5} $\lambda 1238$ & -4 $\pm$ 2 & 11 $\pm$ 5 \\  
                & 1554.9* & \ion{C}{4} $\lambda 1549$ & -2.8 $\pm$ 0.5 & 7 $\pm$ 1 \\   
                & 1751.6  & \ion{N}{3} $\lambda 1751$ & -2.5 $\pm$ 0.6 & 5 $\pm$ 1 \\   
                & 1909.3  & \ion{C}{3} $\lambda 1909$ & -13 $\pm$ 1 & 23 $\pm$ 1 \\  
                & 2468.2  & \ion{O}{2} $\lambda 2470$ & -14 $\pm$ 1 & 15 $\pm$ 1 \\  
\\   
{\bf Hb 7}      & 1245.2* & \ion{N}{5} $\lambda 1238$ & saturated & saturated \\ 
                & 1554.3* & \ion{C}{4} $\lambda 1549$ & -2.7 $\pm$ 0.4 & 10 $\pm$ 1 \\  
                & 1907.9  & \ion{C}{3} $\lambda 1909$ & saturated & saturated \\
\\
\enddata

%% Text for table notes should follow after the \enddata but before
%% the \end{deluxetable}. Make sure there is at least one \tablenotemark
%% in the table for each \tablenotetext.
\tablecomments{An asterisk denotes P-Cygni emission components. Ly$\alpha$ is present in several   
  spectra and has geocoronal origin.}
\end{deluxetable}

%%%%%%%%%%%%%%%%%%%%%%%%%%%%%%%%%%%%%%%%%%%%%%%%%%%%%%%%%%%%%%%%%%%%%%%%%%%%   

\clearpage 
\begin{deluxetable}{lclcc}
\tabletypesize{\small}
\tablecaption{WELS with weak \ion{C}{4} $\lambda 1549$ in P-Cygni.}
\tablewidth{0pt}
\tablehead{
\colhead{{\bf Star}} & \colhead{$\lambda_{obs}$} & \colhead{  Transition } &
\colhead{ W$_{\lambda}$ } & \colhead{ Flux ($10^{-13}$ ergs s$^{-1}$ cm$^{-2}$) }
}

\startdata

{\bf Cn 3-1} &  1397.3* & \ion{Si}{4} $\lambda 1394$ & -0.8 $\pm$ 0.1 & 1.7 $\pm$ 0.2 \\    
             &  1406.5* & \ion{Si}{4} $\lambda 1403$ & -1.1 $\pm$ 0.1 &  2.4 $\pm$ 0.1 \\    
             &  1554.1* & \ion{C}{4}  $\lambda 1549$ & -1.1 $\pm$ 0.6 &  2 $\pm$ 1 \\    
             &  2470.2  & \ion{O}{2}  $\lambda 2470$ & -3 $\pm$ 1     &  4 $\pm$ 1 \\    
\\   
{\bf Hen 2-131} & 1394.7* & \ion{Si}{4} $\lambda 1394$ & -0.8 $\pm$ 0.2 & 22 $\pm$ 4 \\   
                & 1403.4* & \ion{Si}{4} $\lambda 1403$ & -1.0 $\pm$ 0.3 & 30 $\pm$ 7 \\    
                & 1551.6* & \ion{C}{4}  $\lambda 1549$ & -1.2 $\pm$ 0.5 & 28 $\pm$ 9 \\   

\enddata

%% Text for table notes should follow after the \enddata but before
%% the \end{deluxetable}. Make sure there is at least one \tablenotemark
%% in the table for each \tablenotetext.
\tablecomments{An asterisk denotes P-Cygni emission components. Ly$\alpha$ is present in several
  spectra and has geocoronal origin.}
\end{deluxetable}

%%%%%%%%%%%%%%%%%%%%%%%%%%%%%%%%%%%%%%%%%%%%%%%%%%%%%%%%%%%%%%%%%%%%%%%%%%%%%%%%%%%%%%%%%%%%%%%%%%   
   
\clearpage
\begin{deluxetable}{lclcc}
\tabletypesize{\small}
\tablecaption{WELS without \ion{C}{4} $\lambda 1549$ in P-Cygni.}
\tablewidth{0pt}
\tablehead{
\colhead{{\bf Star}} & \colhead{$\lambda_{obs}$} & \colhead{  Transition } &
\colhead{ W$_{\lambda}$ } & \colhead{ Flux ($10^{-13}$ ergs s$^{-1}$ cm$^{-2}$) }
}

\startdata

{\bf IC 4997} &  1305.8 & \ion{O}{1} $\lambda 1305$ & -19 $\pm$ 7 & 7 $\pm$ 1 \\   
              &  1483.9 & \ion{N}{4} $\lambda 1483, 86$ & -10 $\pm$ 7 & 4 $\pm$ 1 \\    
              &  1549.8 & \ion{C}{4} $\lambda 1549$ & -71 $\pm$ 13 & 42 $\pm$ 1   \\    
              &  1642.0 & \ion{He}{2} $\lambda 1640$ & -11 $\pm$ 4  &  7 $\pm$ 1 \\    
              &  1665.0 & \ion{O}{3}  $\lambda 1666$ & -92 $\pm$ 18 &  62 $\pm$ 2 \\    
              &  1750.9 & \ion{N}{3} $\lambda 1750$  & -46 $\pm$ 9  & 30 $\pm$ 2 \\    
              &  1909.1 & \ion{C}{3} $\lambda 1909$  & saturated &  saturated \\   
              &  2324.1 & \ion{O}{3} $\lambda 2321$?, \ion{C}{2} $\lambda 2325$? & -78 $\pm$ 35 & 22 $\pm$ 3 \\   
              &  2473.9 & \ion{O}{2} $\lambda 2470$  & -25 $\pm$ 14 & 14 $\pm$ 5 \\   
              &  2799.7 & \ion{Mg}{2} $\lambda 2796-98$ & -14 $\pm$ 6 &  13 $\pm$ 4 \\   
              &  3187.8 & \ion{He}{1} $\lambda 3188$ & -27 $\pm$ 6 & 20 $\pm$ 2 \\   
\\   
{\bf J900}    &  1547.2 & \ion{C}{4} $\lambda 1549$ & -567 $\pm$ 136 & 50 $\pm$ 1 \\   
              &  1638.5 & \ion{He}{2} $\lambda 1640$& -284 $\pm$ 75  & 19 $\pm$ 1 \\    
              &  1664.0 & \ion{O}{3} $\lambda 1666$ & -62 $\pm$ 14 & 3.2 $\pm$ 0.3 \\    
              &  1906.7 & \ion{C}{3} $\lambda 1909$ & saturated &  saturated \\    
              &  2299.8 & \ion{C}{3} $\lambda 2298$ & -75 $\pm$ 50 & 1.4 $\pm$ 0.3 \\    
              &  2329.3 & \ion{O}{3} $\lambda 2321$?, \ion{C}{2} $\lambda 2325$? & -402 $\pm$ 129 &  8.2 $\pm$ 0.1 \\    
              &  2425.7 & \ion{Ne}{4} $\lambda 2422, 24$ & -86 $\pm$ 13 & 4.8 $\pm$ 0.1 \\    
              &  2473.9 & \ion{O}{2} $\lambda 2470$ & -8 $\pm$ 2 & 0.5 $\pm$ 0.1 \\    
              &  2515.7 & \ion{He}{2} $\lambda 2511$ & -11 $\pm$ 3 & 0.8 $\pm$ 0.1 \\    
              &  2735.6 & \ion{He}{2} $\lambda 2733$ & -13 $\pm$ 1 & 1.5 $\pm$ 0.1 \\    
              &  2802.5 & \ion{Mg}{2} $\lambda 2796-98$ & -9 $\pm$ 3 &  1.1 $\pm$ 0.3 \\    
              &  2838.1 & \ion{O}{3} $\lambda 2836$ & -10 $\pm$ 2 & 1.4 $\pm$ 0.1 \\   
              &  3026.4 & \ion{O}{3} $\lambda 3023-26$ & -7 $\pm$ 1 & 1.2 $\pm$ 0.1 \\   
              &  3048.8 & \ion{O}{3} $\lambda 3046$ & -14 $\pm$ 2 &  2.4 $\pm$ 0.2 \\   
              &  3134.7 & \ion{O}{3} $\lambda 3133$ & -88 $\pm$ 9 &  14 $\pm$ 1 \\   
              &  3206.1 & \ion{He}{2} $\lambda 3203$& -39 $\pm$ 3 &  6.4 $\pm$ 0.2 \\   
\\   
{\bf NGC 5873} &  1478.7 & \ion{N}{4} $\lambda 1483, 86$ & -7 $\pm$ 2 & 1.8 $\pm$ 0.2 \\   
               &  1520.9 & \ion{Si}{2} $\lambda 1527$    & -11 $\pm$ 2 & 2.7 $\pm$ 0.4 \\    
               &  1544.7 & \ion{C}{4} $\lambda 1549$     & -106 $\pm$ 19 & 27 $\pm$ 1 \\    
               &  1635.5 & \ion{He}{2} $\lambda 1640$    & -149 $\pm$ 34 &  29 $\pm$ 1 \\    
               &  1660.2 & \ion{O}{3} $\lambda 1666$     & -18 $\pm$ 4 & 3.2 $\pm$ 0.4 \\    
               &  1903.7 & \ion{C}{3} $\lambda 1909$     & -158 $\pm$ 15 & 30 $\pm$ 1 \\   
\\   
{\bf NGC 6818} &  1238.4 & \ion{N}{5} $\lambda 1238$ & -14 $\pm$ 4 & 7 $\pm$ 1 \\   
               &  1401.4 & \ion{O}{4} $\lambda 1397-1405$ & -20 $\pm$ 3 &  18 $\pm$ 1 \\    
               &  1482.7 & \ion{N}{4} $\lambda 1483, 86$  & -24 $\pm$ 2 &  20 $\pm$ 1 \\    
               &  1547.3 & \ion{C}{4} $\lambda 1549$ & -103 $\pm$ 9 & 81 $\pm$ 1 \\    
               &  1600.8 & \ion{Ne}{4} $\lambda 1601$ & -11 $\pm$ 3 &  8 $\pm$ 1 \\    
               &  1638.8 & \ion{He}{2} $\lambda 1640$ & -258 $\pm$ 26 & 230 $\pm$ 10 \\    
               &  1663.4 & \ion{O}{3} $\lambda 1666$  &  -20 $\pm$ 3 & 19 $\pm$ 2 \\    
               &  1749.0 & \ion{N}{3} $\lambda 1750$ &  -23 $\pm$ 3 & 17 $\pm$ 1 \\   
               &  1906.8 & \ion{C}{3} $\lambda 1909$ & -450 $\pm$ 69 &  320 $\pm$ 10 \\   
               &  2327.0 & \ion{C}{2} $\lambda 2323-28$ & -67 $\pm$ 10 &  14 $\pm$ 1 \\   
               &  2425.9 & \ion{Ne}{4} $\lambda 2422, 24$ & -188 $\pm$ 30 &  43 $\pm$ 1 \\   
               &  2512.8 & \ion{He}{2} $\lambda 2511$ & -16 $\pm$ 3 &  4.1 $\pm$ 0.5 \\   
               &  2734.9 & \ion{He}{2} $\lambda 2733$ & -28 $\pm$ 2 &  8.0 $\pm$ 0.3 \\   
               &  2835.2 & \ion{O}{3} $\lambda 2836$  & -18 $\pm$ 3 &  6.1 $\pm$ 0.6 \\   
               &  3023.8 & \ion{O}{3} $\lambda 3023-26$ &  -9.0 $\pm$ 0.7 &  3.1 $\pm$ 0.2 \\   
               &  3046.9 & \ion{O}{3} $\lambda 3043-3047$ & -28 $\pm$ 4 & 10 $\pm$ 1 \\   
               &  3132.2 & \ion{O}{3} $\lambda 3133$  & -145 $\pm$ 17 & 58 $\pm$ 1 \\   
               &  3203.9 & \ion{He}{2} $\lambda 3203$ & -37 $\pm$ 7 & 16 $\pm$ 1 \\   
\\   
{\bf NGC 6803} &  1641.3 & \ion{He}{2} $\lambda 1640$ & -42 $\pm$ 11 & 1.6 $\pm$ 0.1 \\   
               &  1909.4 & \ion{C}{3} $\lambda 1909$  & -20 $\pm$ 3 &  18 $\pm$ 1 \\   
\\   
{\bf IC 5217}  &  1244.5* & \ion{N}{5} $\lambda 1238$ & -5.7 $\pm$ 0.5 &  2.7 $\pm$ 0.2 \\   
               &  1376.8* & \ion{O}{5} $\lambda 1371$ & -1.0 $\pm$ 0.4 &  0.7 $\pm$ 0.2 \\   
               &  1548.5  & \ion{C}{4} $\lambda 1549$ & -14 $\pm$ 1 &  9.2 $\pm$ 0.2 \\   
               &  1639.6  & \ion{He}{2} $\lambda 1640$ & -15 $\pm$ 1 &  9.7 $\pm$ 0.3 \\   
               &  1663.2  & \ion{O}{3} $\lambda 1666$  & -6.3 $\pm$ 0.6 &  3.7 $\pm$ 0.3 \\   
               &  1751.7  & \ion{N}{3} $\lambda 1750$  & -2.4 $\pm$ 0.4 &  1.4 $\pm$ 0.2 \\   
               &  1906.9  & \ion{C}{3} $\lambda \lambda 1909$ & -54 $\pm$ 4 &  21 $\pm$ 1 \\   
               &  2333.2  &  ?    &  -29 $\pm$ 10 &  5.1 $\pm$ 0.9 \\   
               &  3133.1  & \ion{O}{3} $\lambda 3133$ & -26 $\pm$ 7 & 6.4 $\pm$ 0.9 \\
\\
\enddata

\tablecomments{An asterisk denotes P-Cygni emission components. Ly$\alpha$ is present in several   
  spectra and has geocoronal origin.}   
   
\end{deluxetable}

%%%%%%%%%%%%%%%%%%%%%%%%%%%%%%%%%%%%%%%%%%%%%%%%%%%%%%%%%%%%%%%%%%%%%%%%%%%%%%%%%%%%%%%%   

\clearpage
\begin{deluxetable}{lclcc}
\tabletypesize{\small}
\tablecaption{Line identifications and measurements for A30 and A78.}
\tablewidth{0pt}
\tablehead{
\colhead{{\bf Star}} & \colhead{$\lambda_{obs}$} & \colhead{  Transition } &
\colhead{ W$_{\lambda}$ } & \colhead{ Flux ($10^{-13}$ ergs s$^{-1}$ cm$^{-2}$) }
}

\startdata

{\bf A 30}  &  1246.2* & \ion{N}{5} $\lambda 1238$ & -6.1 $\pm$ 0.8 & 16 $\pm$ 1 \\   
            &  1374.2* & \ion{O}{5} $\lambda 1371$ & -5.0 $\pm$ 0.6 & 13 $\pm$ 1 \\   
            &  1556.7* & \ion{C}{4} $\lambda 1549$ & -9 $\pm$ 1 & 18 $\pm$ 2 \\   
\\   
{\bf A 78}  &  1246.2* & \ion{N}{5} $\lambda 1238$ & -8 $\pm$ 1 & 140 $\pm$ 10 \\   
            &  1374.3* & \ion{O}{5} $\lambda 1371$ & -3.7 $\pm$ 0.5 & 54 $\pm$ 6 \\   
            &  1554.2* & \ion{C}{4} $\lambda 1549$ & -7 $\pm$ 1 & 77 $\pm$ 8 \\   
\\
\enddata

\label{wcpg1159}

\tablecomments{An asterisk denotes P-Cygni emission components. Ly$\alpha$ is present in several   
  spectra and has geocoronal origin.}   
   
\end{deluxetable}

%%%%%%%%%%%%%%%%%%%%%%%%%%%%%%%%%%%%%%%%%%%%%%%%%%%%%%%%%%%%%%%%%%%%%%%%%%%%%%%   

\clearpage
\begin{deluxetable}{lll}
\tablecaption{Terminal velocities of WELS and [WC]-PG 1159 stars.}
\tablewidth{0pt}
\tablehead{
\colhead{Star} & \multicolumn{2}{c}{$v_\infty$ (km\,s$^{-1}$)} \\
\colhead{} & \colhead{\ion{C}{4} $\lambda 1549$} & \colhead{\ion{N}{5} $\lambda 1238$} 
}

\startdata
{\bf WELS}& & \\  
Vy2-1     & 1060        & -           \\   
NGC 6891  & 1420 (1396) & 1200 (1339) \\   
NGC 6629  & 1230        & -           \\   
HEN 2-12  & 1100        & 1230        \\   
PB 8      & 1070        & 1220        \\   
HEN 2-108 & 720         & -           \\   
IC 4776   & 1760        & 1770        \\   
HB 7      & 1077        & 1150        \\   
NGC 6567  & 1750        & 1680        \\   
NGC 6572  & 1070 (1280) & 1200 (1206) \\   
NGC 6543  & 1420 (1684) & 1260 (1552) \\   
CN 3-1*   & 329         & -           \\   
HEN 2-131*& 710 (520)   & -           \\                      
                                                 \\ 
{\bf [WC]-PG 1159}& & \\  
A 30      & 3080 & 2600                   \\ 
A 78      & 2870 (3080) & 2594 (2780)     \\ 
\\
\enddata 
 
\tablecomments{An asterisk indicates stars with weak \ion{C}{4} $\lambda 1549$ 
  in P-Cygni (Group 2). Measurements in high resolution spectra are shown  
between parenthesis.} 
 
\label{vinf}   
\end{deluxetable}

%%%%%%%%%%%%%%%%%%%%%%%%%%%%%%%%%%%%%%%%%%%%%%%%%%%%%%%%%%%%%%%%%%%%%%%%%%%%%%%%%%%%%%%%   

\clearpage
   
\end{document}